\documentclass[conference]{IEEEtran}
\IEEEoverridecommandlockouts
\usepackage{cite}
\usepackage{amsmath,amssymb,amsfonts}
\usepackage{algorithmic}
\usepackage{graphicx}
\usepackage{textcomp}
\usepackage{xcolor}
\usepackage{subfig}
\usepackage{balance}

\newcommand{\MGMark}{MGMark}
\newcommand{\MSGPU}{M-SGPU}
\newcommand{\UMGPU}{U-MGPU}
\newcommand{\DMGPU}{D-MGPU}

\def\BibTeX{{\rm B\kern-.05em{\sc i\kern-.025em b}\kern-.08em
    T\kern-.1667em\lower.7ex\hbox{E}\kern-.125emX}}
\begin{document}

\title{MGSim + \MGMark: A Framework for Multi-GPU System Research}

\author{
    \IEEEauthorblockN{
        Yifan Sun\IEEEauthorrefmark{1}, 
        Trinayan Baruah\IEEEauthorrefmark{1}, 
        Saiful A. Mojumder\IEEEauthorrefmark{2}, 
        Shi Dong\IEEEauthorrefmark{1}, 
        Rafael Ubal\IEEEauthorrefmark{1}, 
        Xiang Gong\IEEEauthorrefmark{1}, \\ 
        Shane Treadway\IEEEauthorrefmark{1}, 
        Yuhui Bao\IEEEauthorrefmark{1}, 
        Vincent Zhao\IEEEauthorrefmark{1},
        Harrison Barclay\IEEEauthorrefmark{1}, 
        Jos\'e L. Abell\'an\IEEEauthorrefmark{3}, \\
        John Kim\IEEEauthorrefmark{4}, 
        Ajay Joshi\IEEEauthorrefmark{2}, and 
        David Kaeli\IEEEauthorrefmark{1}
        \vspace{0.5ex}
    }
    \IEEEauthorblockA{
        \IEEEauthorrefmark{1} Northeastern University\\
        \{yifansun, tbaruah, shidong, ubal, xgong, streadwa, ybao, vzhao, hbarclay, kaeli\}@ece.neu.edu
        \vspace{1ex}
    }
    \IEEEauthorblockA{
        \IEEEauthorrefmark{2} Botson University\\
        \{msam, joshi\}@bu.edu
        \vspace{1ex}
    }
    \IEEEauthorblockA{
        \IEEEauthorrefmark{3} Universidad Cat\'olica San Antonio Murcia\\
        jlabellan@ucam.edu
        \vspace{0.5ex}
    }
    \IEEEauthorblockA{
        \IEEEauthorrefmark{3} KAIST\\
        jjk12@kaist.edu
    }

}


\maketitle

\begin{abstract}

  The rapidly growing popularity and scale of data-parallel workloads demand a
  corresponding increase in raw computational power of GPUs (Graphics
  Processing Units).  As single-GPU systems struggle to satisfy the performance
  demands, multi-GPU systems have begun to dominate the high-performance
  computing world.  The advent of such systems raises a number of design
  challenges, including the GPU microarchitecture, multi-GPU interconnect
  fabrics,
  runtime libraries and associated programming models. 
  The research community currently lacks
  a publically available and comprehensive multi-GPU simulation framework and
  benchmark suite to evaluate multi-GPU system design solutions.

  In this work, we present MGSim, a cycle-accurate, extensively validated,
  multi-GPU simulator, based on AMD's Graphics Core Next 3 (GCN3) instruction
  set architecture.  We complement MGSim with \MGMark, a suite of
  multi-GPU workloads that explores multi-GPU collaborative execution
  patterns. Our simulator is scalable and comes with in-built support for
  multi-threaded execution to enable fast and efficient simulations.  In 
  terms of performance accuracy, MGSim
  differs $5.5\%$ on avarage when compared against actual GPU hardware.
  We also achieve a $3.5\times$ and a $2.5\times$ average speedup in function
  emulation and architectural simulation with 4 CPU cores, while 
  delivering the same accuracy as the serial simulation.
  
  We illustrate the novel simulation capabilities provided by our simulator
  through a case study exploring programming models based on a unified
  multi-GPU system~(\UMGPU) and a discrete multi-GPU system~(\DMGPU) 
  that both utilize unified memory space and cross-GPU memory access. 
  We evaluate the design implications from our case study, suggesting that 
  \DMGPU{} is an attractive programming model for future multi-GPU systems.

\end{abstract}

\section{Introduction}

Given the continued growth of data requirements and processing intensity of a
range of scientific and data-centric applications, a single GPU can no longer
supply the necessary computing power to meet the needs of tomorrow's
applications~\cite{jiang2015scaling, wu2015deep}.  Limited by device technology
challenges of CMOS scaling and associated manufacturing costs, it is becoming
impractical to improve single GPU performance by adding more computing
resources on the same die~\cite{mcmgpu}. 

One potential solution to this issue, which is quickly gaining
popularity, is to integrate multiple GPUs into a single platform. NVIDIA
recently started offering a multi-GPU DGX platform~\cite{dgx}, focusing on
accelerating Deep Neural Network (DNN) training. However, recent studies
suggest that the performance of multi-GPU systems can be heavily constrained by
CPU-to-GPU and GPU-to-GPU synchronization, and limited by multi-GPU memory
management overhead~\cite{umh}. Design of an effective memory management system
and cross-GPU communication fabric remains an open problem that needs to be
addressed to unlock the full potential of future multi-GPU systems.

Recent improvements to GPU memory systems provide support for a {\em unified
memory space}, enabling cross-GPU memory access and system-level atomics.  With
unified memory~\cite{hsa, nvidia_unified_memory} and cross-GPU memory
access~\cite{gpudirect}, one GPU can access the data on another without the
help of the CPU\@.  Additionally, system-level atomics allow for
synchronization across GPUs during the execution of a GPU program.  These
features open up new possibilities for GPU applications to enjoy significantly
improved levels of performance.  At the same time, there are new challenges
associated with multi-GPU systems, including how to handle GPU-to-GPU
communication, memory management across the unified
CPU and multi-GPU memory
space, and application/data partitioning/mapping. These challenges have to be
tackled properly to fully unlock the potential performance of multi-GPU
systems. We need to develop a better understanding of these challenges at a
microarchitectural level.

Due to the lack of appropriate simulation research tools, computer architecture
researchers are handicapped when trying to study the potential of multi-GPU
collaborative execution.  Existing studies~\cite{mcmgpu, umh} on multi-GPU
systems assume a unified logical GPU programming model, which hides the
complexity of the multi-GPU system and exposes only one logical GPU to the
programmer. Despite its simplicity, a unified logical GPU is not necessarily
the right choice for GPU programmers today. The main reason is that on today's
commercial multi-GPU systems, programmers are tasked with precisely controlling
which GPU stores a piece of data and which GPU runs a program, as most
state-of-the-art programming frameworks (e.g., OpenCL~\cite{opencl},
HSA~\cite{hsa}, and CUDA~\cite{cuda}) adopt this model, and major GPU libraries
(e.g., Caffe~\cite{caffe}, TensorFlow~\cite{tensorflow}, etc.) follow this
model.  Researchers cannot use existing GPU simulation frameworks to simulate
multi-GPU systems mainly due to: (1) a lack of modularity and state
encapsulation, and (2) non-scalable performance. 

Due to this lack of modularity and state encapsulation, configuring a multi-GPU
system, or any other major model modification to the simulator,  would require
shotgun-style refactoring. This would involve modifications to major portions
of the simulator codebase, rendering changes to the design a major endeavor.
Further, lack of modularity and state encapsulation inhibits contributions by
the broader research community, given that common files will need modification
by each contributor.  Therefore, the research community is calling for a new
generation of ``modular, pluggable, hookable, and
composable''~\cite{kubernetes} simulators that can provide a much higher level
of extensibility and can support the needs of the user base.

In addition, simulating a large-scale system demands a highly performant
simulator infrastructure.  Using existing simulation frameworks, researchers
can wait for days or weeks to produce a single simulation result. This
simulation cost is further exacerbated when simulating a multi-GPU system, as
simulation overhead will grow faster than linear as we add more simulated
elements.  The major reason is that as the contention in key system components
increases, including interconnects, shared caches, and memory controllers, the
simulation can be several orders of magnitude slower than simulating each GPU
independently. Experimental studies~\cite{lee2013parallel, gputejas} have
explored creating multi-threaded simulators to accelerate simulation. However,
they trade-off accuracy for faster simulation. To provide scalable performance
in a multi-GPU simulation framework, we need a new parallelization approach
that enables system simulation to achieve both high performance and high
accuracy.

In this work, we present MGSim, an open-source
\footnote{https://gitlab.com/akita/gcn3}, cycle-accurate GPU simulator that is
specifically designed for, but not limited to, multi-GPU system simulation.
MGSim executes unmodified Graphics Core Next~(GCN) 3rd generation instruction
set architecture~(ISA) binaries. It is flexible and fully configurable,
enabling users to quickly create a multi-GPU platform for simulation.  MGSim
ships with built-in multi-threading capability, supporting both efficient
functional emulation and architectural simulation, without
compromising simulation accuracy.  

To accompany MGSim, we have developed \MGMark, a new benchmark suite designed
to support design space exploration using multi-GPU {\em collaborative
execution patterns}.  We define multi-GPU collaborative execution as workloads
that leverage multiple GPUs executing the single application, though running
concurrently.  We categorize multi-GPU collaborative execution patterns into
five groups: i) \emph{Partitioned Data}, ii) \emph{Adjacent Access}, iii)
\emph{Gather}, iv) \emph{Scatter}, and v) \emph{Irregular}.  These patterns
span the gamut of general communication schemes that include no data sharing
among the GPUs, to sharing the entire address space for both read and write
operations.  We provide a set of workloads covering not only each multi-GPU
collaborative execution pattern, but also a wide range of algorithms and GPU
features.


MGSim and \MGMark{} form a brand new framework to support the computer
architecture design community.  They allow the community to
efficiently explore a range of multi-GPU models: execution on a discrete
multi-GPU system~(\DMGPU), and execution on a multi-GPU system behind a unified
logical GPU interface~(\UMGPU). To demonstrate this capability, we conduct a
case study that runs \MGMark{} on MGSim and compares \DMGPU{} and \UMGPU,
building a configuration using 4 GPUs in the system.  We explore which GPU
collaborative execution patterns perform well when
targeting each multi-GPU
configuration. 
The main purpose of this study is to demonstrate the the power 
of this new framework, 
while also providing design directions for the future
multi-GPU systems.

The contributions of this paper include the following:

\begin{itemize}

    \item We present a set of design principles that all simulators 
      should aspire to.

    \item We present MGSim, a new parallel cycle-accurate multi-GPU
      architectural simulator that delivers both flexibility 
      and high performance. We extensively validate
      MGSim against real hardware with both micro-benchmarks and
      full workloads.

    \item We present \MGMark, a benchmark suite that explores
        multi-GPU communication patterns.

    \item We use MGSim and \MGMark{} in a case study to analyze the impact of
      unifying  multiple GPUs behind a logical GPU interface. We discuss
      future multi-GPU system design implications from this case study.

\end{itemize}

\section{Background}

\subsection{Multi-GPU Systems}

While today's GPUs are quite powerful, in many emerging applications a single
GPU cannot meet the required processing demands, due to: (1) limited compute
capabilities, and (2) limited memory space.  For example, VGGNet~\cite{vggnet},
a popular deep neural network framework, requires $\approx40$ G-Ops~(Giga
operations) to process a single image through a DNN model~\cite{dnnsize}.
If an
application requires a 1000-images per second throughput (i.e., 40 TFLOPs), we
need, in theory, at least five R9 Nano GPUs to fulfill this requirement.  
On the other hand, training a DNN may require a multi-terabyte
dataset~\cite{imagenet}, dwarfing the GPU memory capacity of a single GPU\@.  If
we can increase the storage available, we have the potential of improving
memory management and significantly accelerating training throughput.
Multi-GPU systems can be a potential solution, as they both provide more compute
resources and more memory storage.

\begin{figure}[!t]
\subfloat[GPU programming model as in
    GPU programming frameworks.\label{fig:discrete_gpus}]{
    \includegraphics[width=0.45\columnwidth]{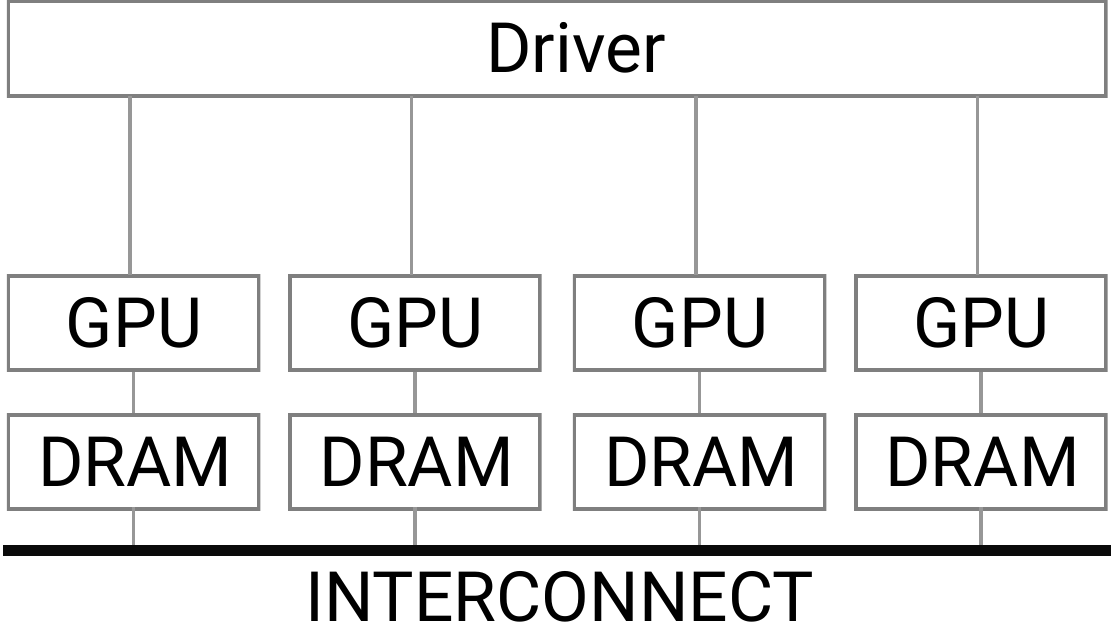}
}
\hfill
\subfloat[GPU programming model as in
    microarchitecture studies.\label{fig:logical_gpus}]{
    \includegraphics[width=0.45\columnwidth]{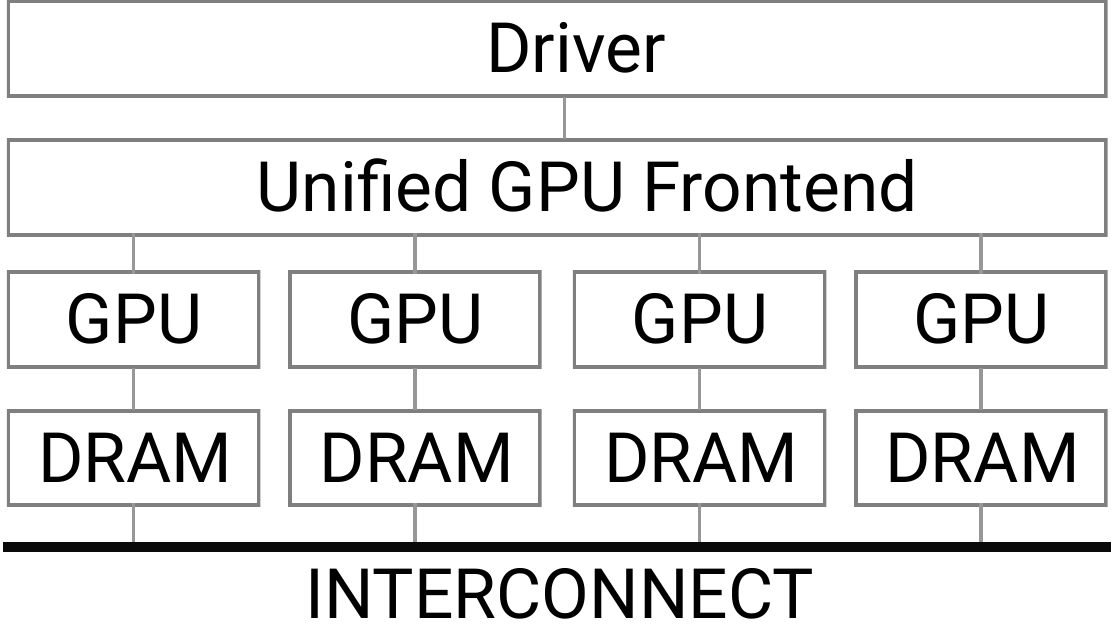}
}
\vspace{-0.1cm}
\caption{Multi-GPU Configuration}
\vspace{-0.1cm}
\end{figure}

The industry has begun to realize the potential of multi-GPU systems \cite{dgx,
kanter2015graphics}. Most popular GPU programming frameworks, including OpenCL
and CUDA, support multi-GPU programming following the model shown in
Figure~\ref{fig:discrete_gpus}.  These programming frameworks expose all of the
GPUs to users, enabling them to select where data is stored and 
how kernels are
mapped to devices.  For example, in OpenCL, a command queue is associated with a
GPU and all the commands (e.g., memory copy, kernel launching) in the queue run
on the associated GPU\@.  In CUDA, the developer can select the GPU to use with
the \texttt{cudaSetDevice} API, and all memory operations and kernel launches
after the API call will target the designated GPU\@.

The computer architecture community lacks simulation frameworks that
can support evaluation of this well-accepted multi-GPU programming model.
Microarchitectural studies usually use a GPU programming model similar to the
one shown in Figure~\ref{fig:logical_gpus}.  Researchers usually simulate
only a single GPU, but configure the interconnect in the simulator to mimic a
multi-GPU system~\cite{umh}.

One feature that greatly simplifies multi-GPU programming is unified
memory~\cite{cuda} (for CUDA) or shared virtual memory~\cite{opencl} (for
OpenCL).  Unified memory provides the programmer with a single unified address
space and avoids the need for explicit memory copies from one device to
another.  Recent unified memory implementations~\cite{gpudirect} even support
cross-GPU memory access, further simplifying the programming model. However, in
current multi-GPU systems, cross-GPU memory access involves data transfers
across a slow interconnect.  Even NVLink, the most recent cross-GPU
interconnect, provides an inter-GPU fabric that is an order magnitude slower
than local memory on the GPU (NVLink supports 20GB/s~\cite{nvlink}, whereas HBM
memory supports 256GB/s~\cite{hbm}).  Therefore, bottlenecks in the cross-GPU
memory system can significantly impact the benefits of multi-GPU systems.  To
begin to design more performant and scalable multi-GPU systems, a new brand of
tools are required to guide design choices.

\subsection{GPU Execution Model}

A typical GPU system is made up of 1 or multiple CPUs, and a few GPUs (up to 8
per node 
in high-end systems).  The GPUs generally are under the control of a CPU\@.
More specifically, the host program that runs on the CPU sets up the data for
the GPU and launches GPU programs (kernels) on GPUs.  A vendor-specific GPU
driver, running at the operating system level, receives requests from the host
program and transfers data and launches kernels on the GPU hardware.

When running on a GPU, a kernel can launch a 1-, 2-, or 3-dimensional grid of
work-items. One work-item is comparable to a thread on a CPU and has its own
register state.  A grid can be divided into work-groups and wavefronts.  On an
AMD GCN3 GPU, a wavefront consists of 64 work-items that execute the same
instruction concurrently.  A work-group contains 1-8 wavefronts which can be
synchronized using barriers.

The design of a GPU supports very high throughput for data-parallel workloads.
For example, the AMD R9 Nano GPU~\cite{r9nano} leverages many Compute
Units~(CU) to execute instructions.  A single CU incorporates four
Single-Instruction Multiple-Data~(SIMD) units.  Each SIMD unit has 32 lanes,
with each lane providing a single precision floating point unit. Hence, a
single SIMD unit can execute 32 instructions in parallel within a single clock 
cycle.
With 64 CUs, the R9 Nano GPU executes up to $64 \times 4 \times 32 = 8,192$
instructions per cycle.  As the R9 Nano GPU runs at a 1GHz clock rate, it can
support a peak throughput of 8,192 $\times$ 1G $=$ 8.19 
terafloating-point operations per second~(TFLOPs).

\subsection{Parallel GPU Simulation}

Architectural simulation can be much slower than
running on real hardware.  For example, Multi2Sim~\cite{multi2sim} is reported
to be $44,000\times$ slower than native execution, 
which translates to more than a
day to simulate 2 seconds of native execution.  Malhotra et,
al.~\cite{gputejas} report that GPGPUsim is $480,000\times$ 
slower than native --- 11
days to simulate 2 seconds of native execution.  These slow simulation speeds
make simulating large-scale systems and large-scale workloads almost impossible
in existing simulators.  To successfully simulate multi-GPUs with large-scale
workloads, we need a new simulation philosophy.


To accelerate architectural simulation, researchers have explored using
multi-threading to accelerate simulation. In general, two types of
parallelization approaches are used: 1) conservative, and 2)
optimistic~\cite{pdes}.  Using a conservative approach, the
chronological order of the events is not interrupted, which requires global
synchronization in each cycle.  
An optimistic approach supports reordering events to avoid frequent 
synchronizations, reducing simulation time, though at some 
cost to the fidelity of the simulation.

\begin{figure}[!t]
    \centering
    \includegraphics[width=\columnwidth]{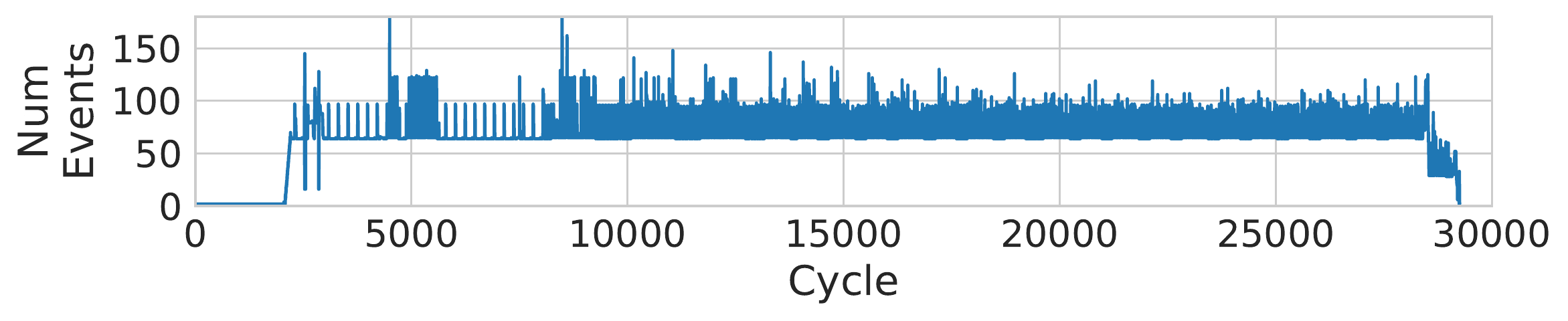}
    \caption{Number of events that can be parallelized 
    without interrupting a chronological order.}
\label{fig:event_count}
\end{figure}

We elect to adopt a conservative parallel simulation scheme because we do not
want to compromise simulation accuracy.  Figure~\ref{fig:event_count} shows the
number of events scheduled at the same time during simulation of the
AES benchmark using MGSim.  We see that the number of events that can be
executed concurrently varies between 60 and 100, providing sufficient
parallelism to keep a 4- to 8-core system busy.

\section{GPU Simulator Design Principles}

Architectural simulators have been one of the most important tools to guide
early design space exploration, performance optimization, and pre-silicon
verification.  Developing an accurate and extensible simulator is essential for
the research community to explore a wide range of design possibilities. 

In the following paragraphs, we discuss a number of design principles 
that simulators should follow, though are absent
in many current simulators.

\textbf{DP-1: Simulate state-of-the-art machine-level ISA.}
Cutting-edge research explores cutting-edge features, and hence,
new ISAs and new microarchitectures need to be evaluated. 
Existing simulators are generally simulating old ISAs or intermediate 
representations. 
For example, Multi2Sim~\cite{multi2sim} emulates the GCN1 ISA, which is 
four generations older than current AMD product. 
GPGPU-Sim~\cite{gpgpusim} mainly models the NVidia Fermi architecture 
that was released in 2010.
In addition, researchers have highlighted major issues 
when performing performance analysis while simulating at an
intermediate language level versus using the actual machine code 
ISA~\cite{multi2sim_kepler, amdgpu}, resulting in misleading performance.  
Therefore, while any simulator will immediately become quickly
dated due to the pace of development in GPU technology,
the research community needs a simulator 
that can simulate a new and feature-rich machine-level ISA\@.

\textbf{DP-2: ``Open to Extension, Closed to
Modification.''} 
When studying performance/power/reliability with an architectural simulator,
researchers usually need to reconfigure, or more commonly, modify, the
simulator to fit the needs of their intended study. Modifying the
inter-dependent components in a simulator is non-trivial and may require
modifying a large number of files. It tends to be more problematic when
combining the modifications from different developers, as each developer may
need to modify common files.

According to the ``Open-Closed Principle''~\cite{martin2002agile}, one should
be able to extend a simulator without modifying it. When adding more
functionality to a simulator, researchers should not need to modify source
files. Instead, they should write new extensions for the simulator and plug the
new extensions into the existing simulator to realize new configurations.
This approach can also help support the reproducibility of results, since
each module can be clearly defined and reused~\cite{reproducible}.

\textbf{DP-3: No magic.} It is tempting for simulator developers to overuse the
flexibility that a software design offers to overcome the complexity of the
simulated hardware design, typically manifested in intricate queuing systems,
asynchronous buffers, or low-level communication protocols. 

As an example of ``magic'', the implementation of a GPU may directly invalidate
the caches by invalidating all directory entries, ignoring the fact that in
real hardware, this action involves a message to be sent from the command
processor to each cache module.  Manipulating the state of one module from
another is a clear sign that the simulator is not tracking the behavior of real
hardware, and this may impact simulation accuracy.  When a simulator developer
uses ``magic'', it hurts both the accuracy of the simulator,
as well as encapsulation and
modularity of the code.

\textbf{DP-4: Track both timing and data.}
Directly inferred from the ``no-magic'' rule, a simulator should model the
actual data-flow in both the memory system and the instruction pipelines,
rather than only calculating the simulated time.  Execution simulation that
maintains data values offers two advantages: (1) Minor mistakes in the
simulator will be detected as a mismatch of output values, rather than a
difference in the estimated time. If the result generated by the simulator
matches execution on the target hardware, we can guarantee that the modeled
hardware is at least feasible. (2) A performance model or power model may be
data dependent~\cite{cmos_data_dependent, nacre}. Maintaining data in each
module under simulation helps us support data-dependent modeling, which can
improve accuracy.

\textbf{DP-5: Simulate multi-threaded hardware with multi-threaded software.} 
A GPU supports a massively parallel execution model. There are a large number
of units concurrently executing independently on a GPU\@. Therefore, it should
be possible to use multiple CPU threads to simulate GPU execution. In addition,
properly applying locks in a multi-threaded program to prevent race conditions
and avoid deadlocks is usually a difficult job.  The design of the simulator
should provide a locking scheme that both guarantees performance and avoids the
hazards described above.

\textbf{DP-6: No busy ticking.} 
Busy ticking (i.e., constant checks of module states) is a common reason for
low simulation performance, and should be avoided.  In current simulator designs
(e.g., GPGPUSim), modules usually need to check their internal state every
cycle, even if the states do not need to be updated. This is a common problem
for cycle-based simulation. Multi2Sim~\cite{multi2sim} partially solves the
problem by using a hybrid cycle-based and event-driven simulation scheme.
However, some modules still need to keep retrying actions each cycle, such as
cache reads to the cache while the network is busy.  To achieve good simulation
performance, a next-generation simulator should avoid busy-ticking whenever
possible.

\section{MGSim}

MGSim is a highly-configurable GPU simulator that is open-sourced under the
terms of the MIT licence~\cite{mit_licence}. The simulator has been developed
using the Go programming language~\cite{EffectiveGo}. We selected Go because
it provides both reasonable performance and ease of programmability. It also
provides native language-level support for multi-threaded programming.

\subsection{Simulator Core}

The simulator core features a lightweight design composed of four parts:

\textbf{1. The event system:} An event marks an update of the system state that
occurs at a particular time. An event-driven simulation engine maintains a queue
of events and triggers events in chronological order.

\textbf{2 The component system:} Every entity that the simulator simulates is a
component. In our case, a GPU, a compute-unit, and a cache controller are
examples of components. A component can only schedule events to itself and
cannot decide what other components do in the future. Each component
serves as an event handler that can process different types of events.
The same type of event may have different behavior when handled 
by different components.

\textbf{3. The request-connection system:} Two components can, and only can, 
communicate with each other through connections using requests.
Connections are used to model the network-on-chip~(NOC) and cross-chip
interconnects.

\textbf{4. The hook system:} Hooks are small pieces of software
that can be configured to attach to the simulator to either read simulation
state, or update the simulator state. The event-driven simulation engine, all
the components, and the connections are hookable. Hooks are used to perform
non-critical tasks such as collecting traces, dumping debugging information,
calculating performance metrics, recording stall reasons, and injecting faults
(for reliability studies).

The MGSim event engine supports parallel simulation, 
fulfilling \emph{DP-5}. 
Leveraging the fact that
the events that are scheduled concurrently do not depend on each other, the
event engine employs multiple CPU threads to process the events that are
scheduled concurrently. We embrace a conservative parallel event-driven scheme,
so that we guarantee accurate results that match execution on a serial version
of the simulator.

The component system and the request-connection system enforce a very
strict state encapsulation of components. Since we restrict a component from
scheduling events for other components, nor allow a component to access
another component's state (by reading/writing field values, using
getter/setter functions,
or function calls), all communication must use the connection systems.  This
design forces the developer to explicitly declare protocols between components.
The benefits of this design are three-fold.

First, a developer can implement a component without any concern for the
communication protocol, letting the request-connection system
worry about the implementation details of
connecting components.

Second, we gain flexibility by allowing the user to compose two components that
follow the same protocol freely. When a researcher wants to extend the
simulator, one does not need to modify the existing simulator, but only needs
to rewrite a new component that replaces an existing component. The researcher
only needs to be compliant with the protocol of the original component.  When
combining efforts of two researchers, one simply needs to import the code from
two sources and write a new configuration to connect the systems together.  By
adopting this model, we fulfill the requirement of \emph{DP-2}. We encourage
researchers that use MGSim to create a new git repository (open-source,
ideally) that only contains the extensions to MGSim, and provides 
the necessary configuration
code to wire the new extension to the original MGSim.

Third, we can improve simulation accuracy as no information can ``magically''
flow from one component to another, without being explicitly transferred
through the interconnect.  As a consequence, the processor cannot access the
data directly in DRAM, forcing all the data to flow through the cache
hierarchy. We do not support emulation run-ahead during architectural
simulation, as the processor does not even have the instruction bytes until the
data is explicitly fetched from memory system. Therefore, we can both satisfy
\emph{DP-3} and \emph{DP-4} with this design. 

The component system also contributes to addressing \emph{DP-5}, by creating a
clear boundary on where locks should be used. As the \texttt{Handle} function
is the only place where a component can update its internal state (other
components cannot access this state), we simply set a lock at the beginning of
the \texttt{Handle} function and unlock at the end of the \texttt{Handle}
function.

The event-driven simulation and the connection system can help avoid
busy ticking (\emph{DP-6}). For long latency actions, such as a 300-cycle
latency in reading DRAM, we can schedule an event in the event-driven
simulation engine after 300
cycles and skip state updates in
between.  In addition, another type of busy ticking in GPU architectures is
caused by components that repeatedly retry to send data.  
Since a component has no information about
when a connected connection becomes available, the component has to retry 
each cycle.  To avoid this type of busy ticking, we allow the connections to
explicitly notify connected components when the connection is
available. Therefore, a component can avoid updating the state if all of the
out-going connections are busy, as no progress can be made, and continue to
update cycle-by-cycle after the connection is available.

\subsection{GPU Architecture Simulation}

\begin{figure}[!t] 
\centering 
\includegraphics[width=\columnwidth]{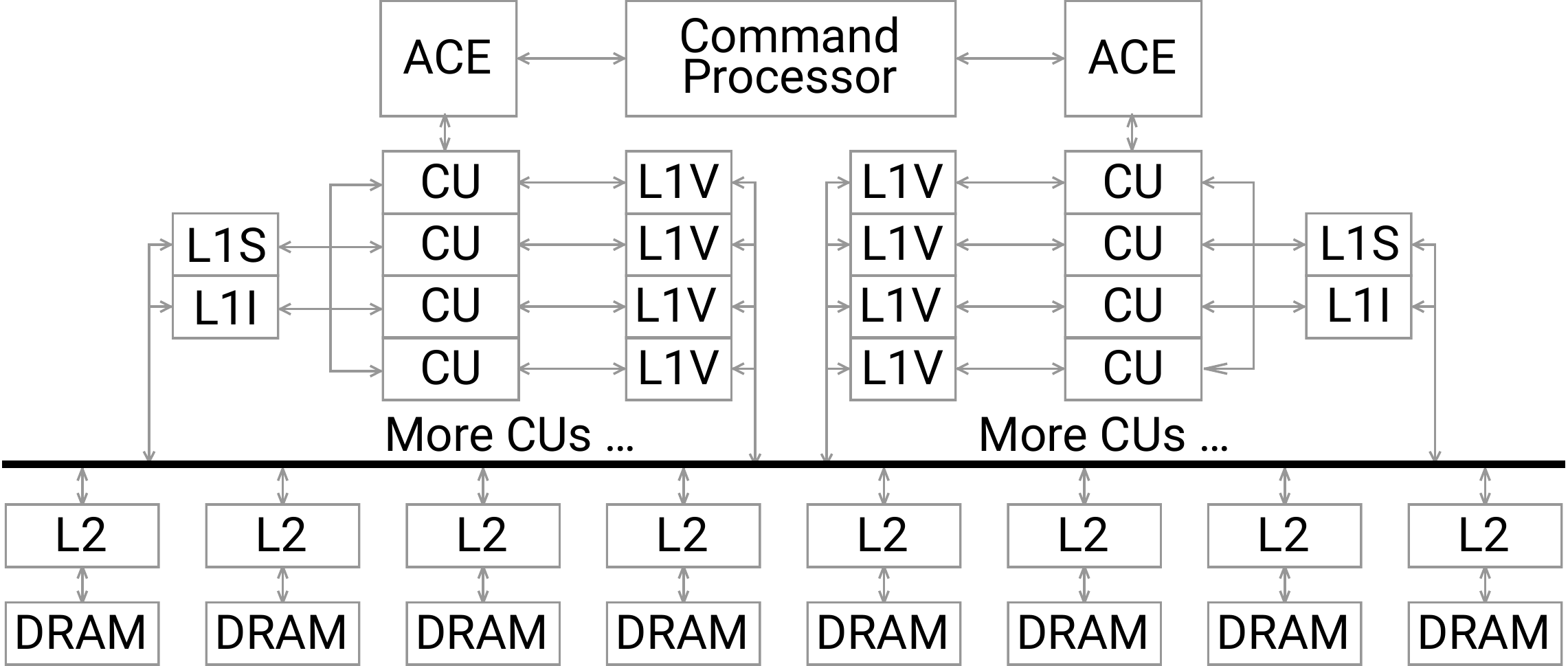}
\caption{The Modeled GPU Architecture. Each block in the figure is a
\emph{Component}.} 
\label{fig:gpu_arch} 
\end{figure}

MGSim models a GPU, as shown in Figure~\ref{fig:gpu_arch}, that runs the
Graphics Core Next 3rd Generation~(GCN3) ISA, fulfilling \emph{DP-1}, 
by simulating a new ISA and microarchitecture. GCN4~\cite{gcn4} and
Vega~(GCN5)~\cite{gcn5} only involve microarchitecture modifications or minor
memory access instruction extensions, and hence, can still be modeled by
configuring MGSim.

\begin{figure}[!t] 
\centering 
\includegraphics[width=\columnwidth]{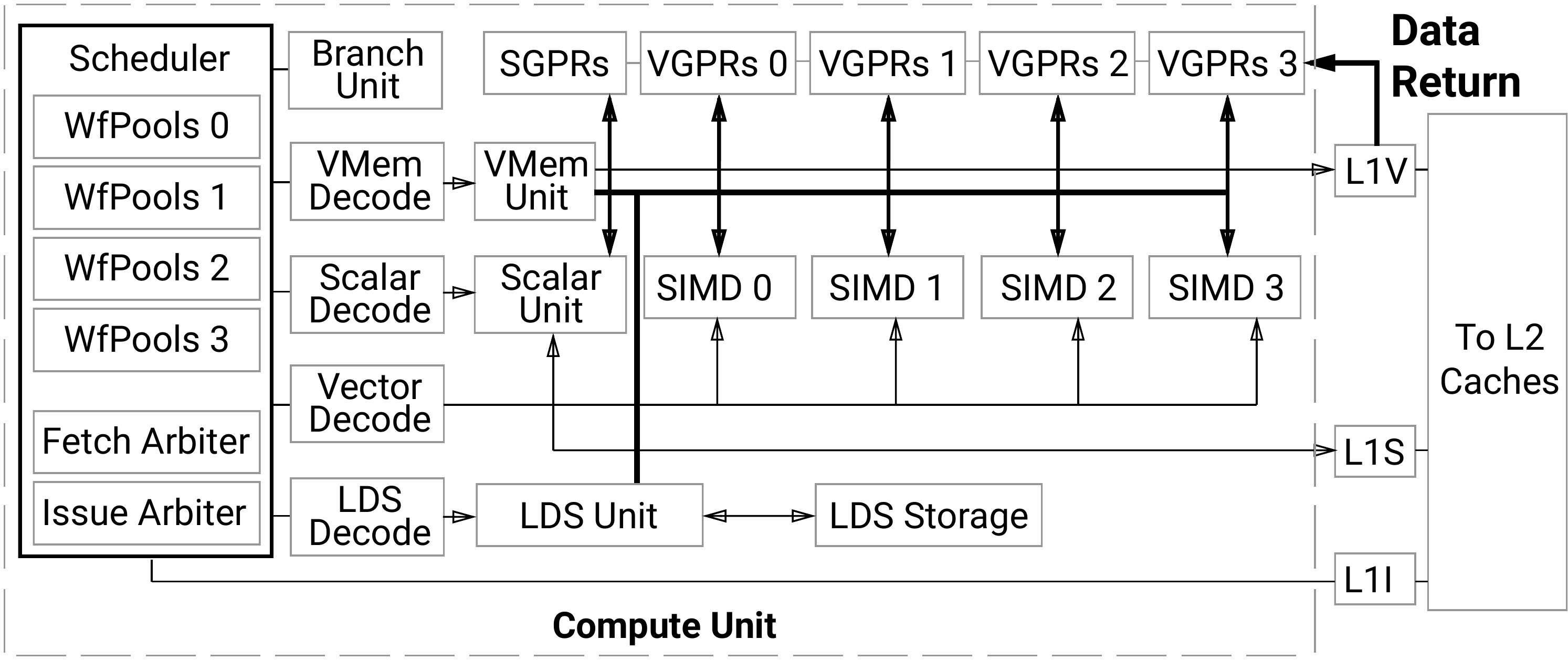}
\caption{The Compute Unit Model} 
\label{fig:cu} 
\end{figure}

The GPU architecture is mainly composed of a Command Processor~(CP),
Asynchronous Compute Engines~(ACEs), Compute Units~(CUs), caches, and memory
controllers. The CP is responsible for communicating with the GPU driver and
starting kernels with the help of ACEs.  The ACEs dispatch wavefronts of
kernels to run on the CUs.

In our model, a CU (as shown in Figure~\ref{fig:cu}) incorporates a scheduler,
a set of decoders, a set of execution units, and a set of storage units. The CU
includes a scalar register file, vector register files, and a local data share
(LDS) storage.  A fetch arbiter and an issue arbiter decide which wavefront can
fetch instructions and issue instructions, respectively, in a round-robin
manner. Decoders require 1 cycle to decode one instruction, before sending the
instruction to the execution unit (e.g., SIMD unit).  Each execution unit has 
a pipelined design that includes read, execute, and write stages. After one
instruction completes all the stages in the pipeline, the wavefront that owns
the instruction can issue the next instruction.

The MGSim simulator includes a set of cache controllers, including a
write-around cache, a write-back cache, and a memory controller.  By default,
the L1 caches and the L2 caches use a write-around and write-back policy,
respectively. The cache controllers do not enforce coherence, as allowed by
the GPU programming and memory model.


\subsection{Multi-GPU Configuration}
\label{subsec:multigpu-confs}

\begin{figure}[!t]
  \subfloat[Unified multi-GPU system~(\UMGPU)\label{fig:unified_gpus}]{
    \includegraphics[width=\columnwidth]{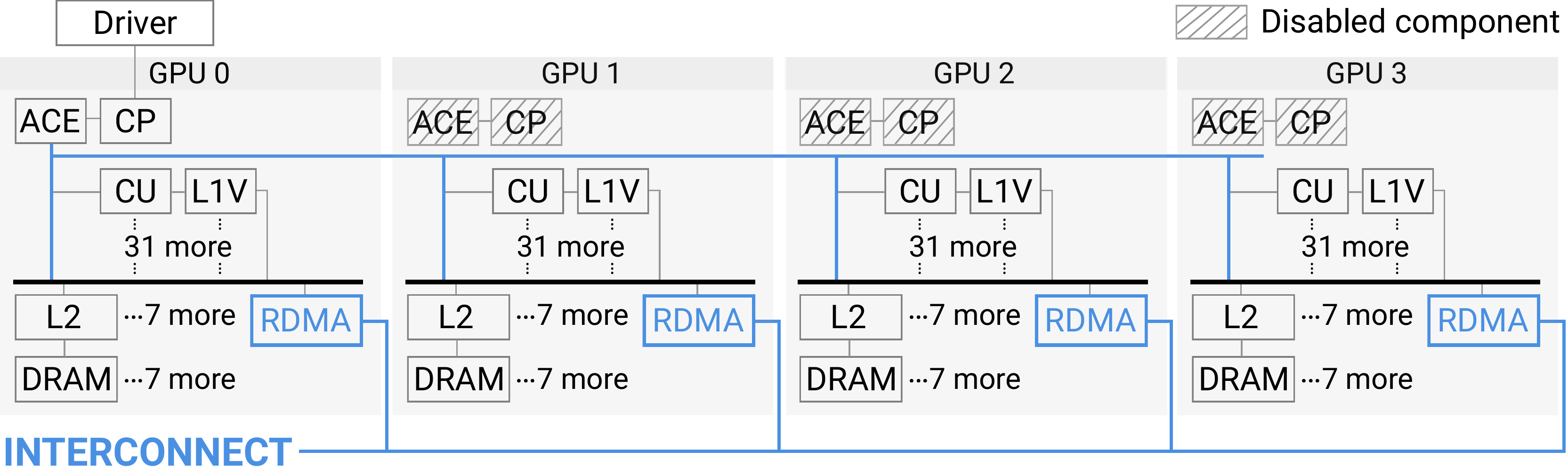}
  }
  \vspace{-0.1cm}
\\
  \subfloat[Discrete multi-GPU system~(\DMGPU)\label{fig:separated_gpus}]{
    \includegraphics[width=\columnwidth]{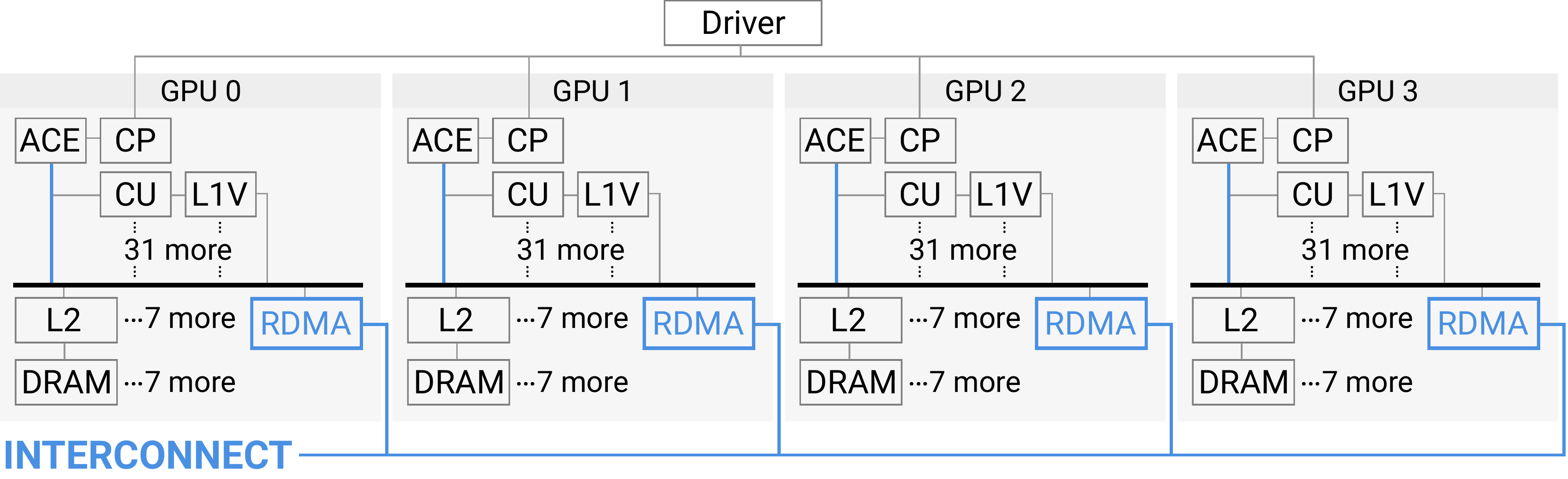}
  }
\caption{Multi-GPU configurations}
\vspace{-0.2cm}
\end{figure}

To demonstrate configurability of the simulator, we explore the multi-GPU
design space, configuring three different multi-GPU platforms --- namely, a
Monolithic Single GPU~(\MSGPU), a Unified multi-GPU system~(\UMGPU), and a
discrete multi-GPU system~(\DMGPU).  The \MSGPU{} is similar to the
base-line R9 Nano GPU configuration (as in Figure~\ref{fig:gpu_arch}), but
provides 256 CUs, 32 L2 cache units, and 32 memory banks, making the computing
power equivalent to four GPUs.  Note that the \MSGPU{} is just a baseline
design to help us analyze performance scaling of the multi-GPU systems.  In
reality, manufacturing such a GPU is impractical due to the limitations of
current die sizes.

As shown in Figure~\ref{fig:separated_gpus}, the \DMGPU{} design creates a GPU
configuration that is commonly provided on current platforms: the driver
connects to multiple GPUs, and the programmer can use APIs to control where the
data resides and where the kernel executes.  To enable unified memory and
cross-GPU memory access, we introduce \emph{RDMA engines} that route
memory requests to other GPUs.  We connect the RDMA engines via a PCIe
bus, providing a bandwidth of 16 GB/s shared by all the GPUs.

To create \UMGPU,  we disable the Command Processors and the ACE of GPUs 1 to
3, leaving the CP and the ACE of GPU 0 in charge of all the Compute Units.  We
also create a cross-GPU connection that connects the ACE of the first GPU with
the CUs of all other GPUs.

The DRAM banks in the multi-GPU systems are interleaved with a granularity of
4KB\@.  For example, the address \texttt{0x0000}-\texttt{0x0FFF} is stored in
DRAM 0 of GPU 0 and so on, 
An exception is in \DMGPU, where the
address space is first partitioned across GPUs and then interleaved, making the
DRAMs of the second GPU map to the address range 4GB-8GB\@.

\section{\MGMark}

\emph{\MGMark} is a new benchmark suite that targets  exploration of multi-GPU
collaborative execution patterns with a wide range of multi-GPU workloads.

\subsection{Multi-GPU Collaborative Execution Patterns}

Execution patterns include types of behavior that repeatedly appear in program
execution. The pattern of a program is usually determined by both algorithm
constraints and implementation decisions. 
In this work, we consider a scenario
where the data to be processed is large, so that duplicating the data to each
GPU adds too much overhead, and is impossible to run on a single GPU due to
memory size limitations.

Studying multi-GPU collaborative execution patterns can help us cover most
types of multi-GPU execution with a smaller number of benchmarks.  It can also
guide programmers and system designers to optimize programs and systems for
specific targets.  Note that the patterns introduced here are not meant to be
exhaustive, nor mutually exclusive.  One multi-GPU program may use more than
one pattern, or may use patterns that we do not characterize in this
paper.

\textbf{Partitioned Data}: The \emph{Partitioned Data} pattern describes a type
of algorithm that naturally allows both the input and output data to be
partitioned on each GPU\@. The result is that no cross-GPU memory accesses are
required.  This pattern is frequently observed in streaming applications, such
as AES encryption~\cite{aes}, and the Blacksholes algorithm~\cite{blackscholes},
where the input and output have a one-to-one mapping.  
This pattern usually relies on a head node (a CPU or a GPU) to partition the
data and broadcast the data to each GPU  to process each batch.  As no
cross-GPU communication is required, this pattern is likely to achieve good
scalability, and hence, should be used whenever possible.

\textbf{Adjacent Access}: The \emph{Adjacent Access} describes a pattern where
the GPUs need to access data, that is closely related to their own local data,
from other GPUs.  This pattern is frequently observed in signal
processing~\cite{fir}, stencil algorithms~\cite{stencil}, and physical
simulations~\cite{hotspot, shallowwater}, as calculating one output at one
particular index needs the input data from surrounding indices that are
resident on a neighboring GPU\@.  If the data that needs to be accessed from
another GPU is read-only, we can maintain multiple copies of the data to avoid
cross-GPU access, at the cost of using more GPU memory space.  Otherwise, we
can keep the data partitioned on each GPU and allow each GPU to issue cross-GPU
accesses occasionally. Adjacent accesses involve a relatively small amount of
cross-GPU communication, and therefore, can be a good option
compared to data duplication.

\textbf{Gather}: This pattern describes a commonly used computing paradigm,
where every GPU in the system needs to 
read remote data from the other GPUs, 
but each GPU will only write to its own local memory.  The \emph{Gather} pattern
can be used in reduction style computing (e.g., adding two vectors element-wise
or calculating the sum of a vector) as each GPU needs to synthesize a larger
amount data to create a smaller output.  When the data is too large to fit in
one GPU's memory, or the data is already on each GPU, we can use a Gather
operation.  The Gather pattern requires the system to process cross-GPU read
requests with rather low latency.

\textbf{Scatter}: Opposite but similar to Gather, 
\emph{Scatter} describes a pattern
where each GPU needs to input data from a local GPU and output data to the
entire GPU address space.  This pattern is used when the input data can be
partitioned on each GPU, while the output location is non-deterministic.

\textbf{Irregular}: We summarize all other patterns as following 
an \emph{Irregular}
pattern, and includes patterns when any GPU needs to both read and write data
from/to the entire GPU address space.  This data reference pattern occurs in
many sorting and graph algorithms, as the access pattern is
data-dependent.  The Irregular pattern presents performance challenges since it
may result in frequent cross-GPU communication. The programmer should try to
use other patterns before settling for an Irregular pattern.  Also, whenever
this pattern is used, the programmer should make every effort to keep memory
accesses within a local GPU and avoid cross-GPU accesses.

\subsection{Workloads}

We select a suite of workloads from public-domain libraries and benchmark
suites, including the AMDAPPSDK 3.0~\cite{amdappsdk}~(BS, MT, SC) and
HeteroMark~\cite{heteromark}~(AES, FIR, KM), as well as one benchmark~(GD)
developed from scratch.  Workloads are modified with new OpenCL kernels
supporting multi-GPU execution, and extended with a Go main program 
compatible with the simulator.

\textbf{Advanced Encryption Standard~(AES)}: 
AES 256-bit encryption~\cite{aes} is an encryption algorithm widely used in the
security domain today.  It involves a large number of bitwise operations to
convert the plaintext to ciphertext, making it a compute-intensive workload.
Our partitioned implementation breaks up the plaintext into chunks and broadcasts
the chunks to the GPUs.  Each GPU then works on its own chunk of the data, with
no need to access any remote data.

We include this benchmark to test the Partitioned Data pattern.  We also use
this benchmark to validate our model for sub-dword-addressing, a distinct
feature of the GCN3 and later AMD ISAs~\cite{gcn3}. 

\textbf{Bitonic Sort~(BS)}: 
Bitonic Sort~\cite{bitonicsort} is a sorting algorithm that suits the GPU's
massively parallel architecture. It has a predefined order to compare pairs of
values in the array to be sorted, making it highly data-parallel.

We include the Bitonic Sort algorithm to test the Irregular pattern.  Although
the memory access order is predefined, each GPU needs to read from, and write
to, any location in the unified memory address space. It also scans a wide
range of memory addresses repetitively, putting significant stress on the cache
system.

\textbf{Finite Impulse Response Filter~(FIR)}: 
FIR~\cite{fir} is a fundamental algorithm from the digital signal processing
domain. In FIR, each work-item multiplies the filter kernel with a portion of
the input data in an element-wise manner and sums all the results together.

We include FIR to test the Adjacent Access pattern, as for each GPU, the first
few work-items on each GPU need to access the input data that is stored on
another GPU\@.  Its large memory footprint can help us analyze how cross-GPU
memory access may have a significant performance impact.

\textbf{Gradient Descent~(GD)}: 
Gradient descent~\cite{gradientdescent} is an important step used in
optimization problems such as DNN Training.  Gradient descent evaluates the
gradient values for a set of mathematical functions and uses the gradient value
to update each function's parameters. When running on a multi-GPU system, gradient
descent is usually performed in a data-parallel fashion, as each GPU processes
a mini-batch of the data (i.e., the Partitioned Data pattern).  At the end of
calculating the gradient on each GPU, the gradient values need to be averaged.
Calculating the average inevitably involves cross-GPU communication.

We include the GD workload as it is one of the most widely used algorithms that
requires the Gather pattern. Its large memory footprint is also a good test case
to stress the cross-GPU interconnect.

\textbf{KMeans~(KM)}: 
KMeans~\cite{kmeans} is an important clustering algorithm widely used in
unsupervised machine learning applications. The GPU is responsible for
calculating the distance from each input node to each of the centroids, while
the CPU updates the centroid location.

We select the KMeans benchmark to evaluate the Partitioned Data pattern.  This
workload is different from AES, which also follows the Partitioned Data
pattern, in two respects: i) KMeans is a more memory intensive workload, and
ii) KMeans repetitively accesses the same memory locations in multiple kernels,
making it more sensitive to the cache design. 

\textbf{Matrix Transpose~(MT)}: 
Matrix Transpose is one of the building blocks common in more complex matrix
operations. Work-items from one work-group first load matrix data to the local
data share memory (an addressable memory space with similar latency to the L1
caches), and then write the data back to the memory in the transposed
locations.

Although MT can be implemented using both the Gather pattern and the Scatter
pattern, we include the Matrix Transpose benchmark to test the Scatter pattern.
Each GPU is responsible for a specific number of columns in the output matrix.
Since each GPU stores a few rows of both the input and output matrix, each GPU
can read from local memory and write to other GPUs.  We also use MT to
validate the simulator on Local Data Store (LDS) operations.

\textbf{Simple Convolution~(SC)}: 
Simple convolution is a common operation in the image processing domain. It is
also a fundamental step in convolutional neural networks~(CNNs). SC
performs a convolution operation on 2-dimensional images. 

We include SC to test the Adjacent Access pattern in a 2-dimensional problem.
Although the image to be convolved can be partitioned across multiple GPUs,
each GPU needs to access a remote partition for the input pixels on
the margins.

\section{Evaluation Methodology}

We start by using both micro-benchmarks and full benchmarks 
from   
\MGMark{} to validate the 
accuracy of MGSim, and then evaluate the 
multi-GPU design performance as a case study.

\subsection{Execution Platforms}

\begin{table}[h!]
  \centering
  \begin{tabular}{|c|c|}
    \hline
    \textbf{Parameter} & \textbf{Value}\\
    \hline
    \hline
    Number of CUs & 64 \\
    \hline
    Core Frequency & 1.0 GHz \\
    \hline
    Theoretical Compute Speed & 8.19 TFLOPs \\
    \hline
    L1 Vector Cache & 64 $\times$ 16KB 4-way\\
    \hline
    L1 Instruction Cache & 16 $\times$ 32KB 4-way\\
    \hline
    L1 Constant Cache Size & 16 $\times$ 16KB 4-way\\
    \hline
    L2 Cache Size & 8 $\times$ 256KB 16-way\\
    \hline
    DRAM Size & 8 $\times$ 512MB \\
    \hline
  \end{tabular}
  \caption{Specifications of the R9 Nano GPU.}
  \vspace{-0.3cm}
  \label{table:r9nano}
\end{table}

In order to validate MGSim against a real hardware, we
collect the actual GPU execution time as a golden performance reference.
The validation system
has 2 Intel Xeon E2560 v4 CPUs and one AMD R9 Nano GPU 
(details provided in Table~\ref{table:r9nano}).
The system
runs the Radeon Open Compute Platform (ROCm) 1.7 GPU software stack 
on a Linux Ubuntu 16.04.4 operating system.
We lock the GPUs to run at the maximum frequency to avoid the 
impact of the Dynamic Frequency and Voltage Scaling~(DVFS) on the system.
All the timing results are collected using the Radeon Compute 
Profiler~\cite{rcp}.

MGSim supports ROCm standard~\cite{rocm}, 
so we compile the benchmarks with AMD's ROCm compiler.
We use \texttt{clang-ocl} to compile the 
\MGMark{} workloads
and use Clang (ROCm modified) to 
assemble the kernels of the micro-benchmarks (to be introduced 
in~\ref{subsec:microbenchmark}).
The host programs are compiled with GCC 5.2.

We evaluate simulation speed and multi-threaded scalability on a 
host platform based on an Intel Core i7-4770 CPU, 
with 4 cores and 2 threads per core.
When measuring the simulator performance, we use the environment 
variable \texttt{GOMAXPROCS} to set 
the number of CPU cores that the simulator can use.

\subsection{Micro-benchmarks}\label{subsec:microbenchmark}

We use a set of micro-benchmarks to help us confirm that our simulator 
can faithfully model each individual aspect of the real hardware. 
Each micro-benchmark is composed of a manually written 
GCN3 assembly kernel, 
a C++ host program used in native execution, 
and an additional host program written in Go for simulation purposes. 
Micro-benchmarks include:


\textbf{ALU}: A simple Python script is used to generate 
kernels with a varying number of ALU operations
(\texttt{v\_add\_f32 v3, v2, v1}), followed by an 
\texttt{s\_endpgm} instruction. 
Using the ALU micro-benchmark, we validate
instruction cache policies and geometry.

\textbf{L1 Access}: Another Python program is again used 
to generate a fixed number of memory reads to the same address. 
All accesses, except for the first one, are presumably L1 cache hits, 
which allows us to infer the cache latency.

\textbf{DRAM Access}: Global memory is repeatedly accessed 
using a 64-byte stride. Since all cache levels use 64-byte blocks, 
all accesses are expected to incur cache misses, 
and ultimately read from the DRAM\@. 
We use this micro-benchmark to measure the DRAM latency.

\textbf{L2 Access}: This micro-benchmark first scans 
1MB of memory, loading all of the data in the 
2MB L2 cache on the R9 Nano.
The L1 cache is expected to retain the last 16KB, 
which is equal in size to its total capacity. After this, a second scan 
sweeps the same 1MB of data from the beginning, 
using a variable number of memory accesses. 
All the memory accesses in the second pass 
should miss in the L1 cache and hit in L2. 
We use this strategy to find the L2 cache latency.

\subsection{\MGMark{} Configuration}

\begin{table}[h!]
  \centering
  \vspace{-0.1cm}
  \begin{tabular}{|c|c|c|c|}
    \hline
     & \textbf{1 GPU} & \textbf{4 GPUs} & \textbf{What}\\
    \hline
    \hline
    AES & 256KB & 1MB & Plaintext\\
    \hline
    BS & 32K & 128K & SP numbers\\
    \hline
    FIR & 64K & 256K & SP samples\\
    \hline
    GD & 256K & 1M & SP parameters\\
    \hline
    KM & 32K & 128K & 32-feature points\\
    \hline
    MT & 2048 & 4096 & Width of square matrix\\
    \hline
    SC & 1024 & 2048 & Width of square image\\
    \hline
  \end{tabular}
  \caption{\MGMark{} configuration for single-GPU validation experiment 
  and multi-GPU case study.}
  \vspace{-0.2cm}
  \label{table:benchmark}
\end{table}

We configure the benchmarks as specified in the ``1 GPU''
column of Table~\ref{table:benchmark} to further 
validate MGSim with \MGMark.
To stress the multi-GPU system with heavier workloads, 
we quadruple the workload size in the multi-GPU 
experiments, to be discussed further in~\ref{sec:multi-gpu-result}.

\section{Experimental Results}

We first carry out a thorough experimental evaluation
to validate our simulator against GPU hardware. 
Then we present a set of experiments to demonstrate 
how microarchitecture design can impact 
multi-GPU collaborative execution efficiency.

\subsection{Simulator Validation with Micro-benchmarks}





\begin{figure}[!t]
\subfloat[ALU Instructions\vspace{-0.3cm}\label{fig:alu_micro_benchmark}] {
    \includegraphics[width=\columnwidth]{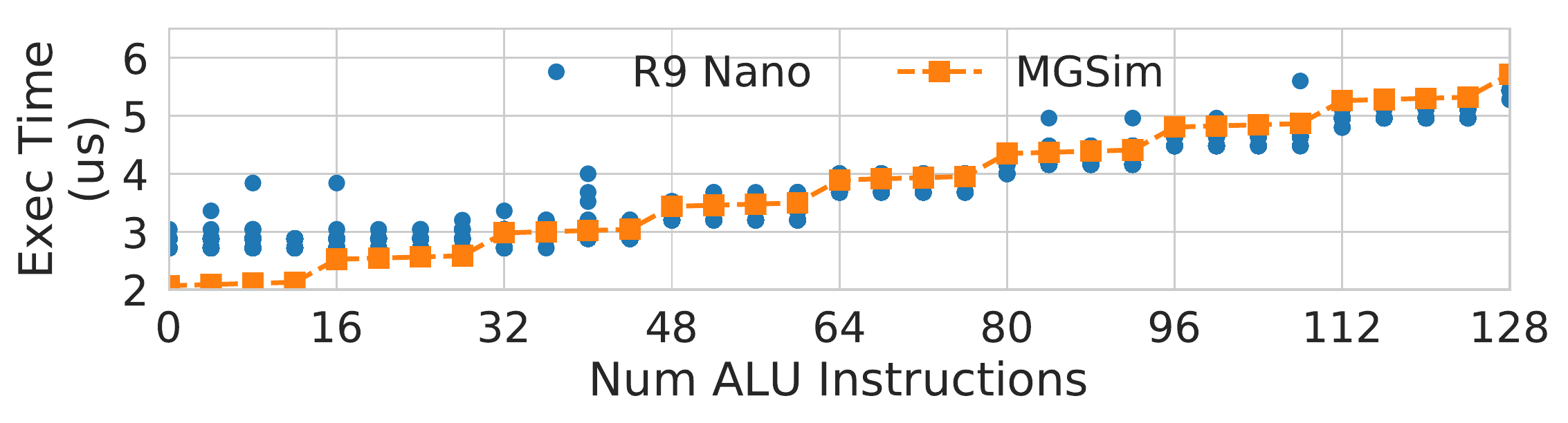}
}\\
\subfloat[L1 Read\vspace{-0.3cm}\label{fig:l1_micro_benchmark}] {
    \includegraphics[width=\columnwidth]{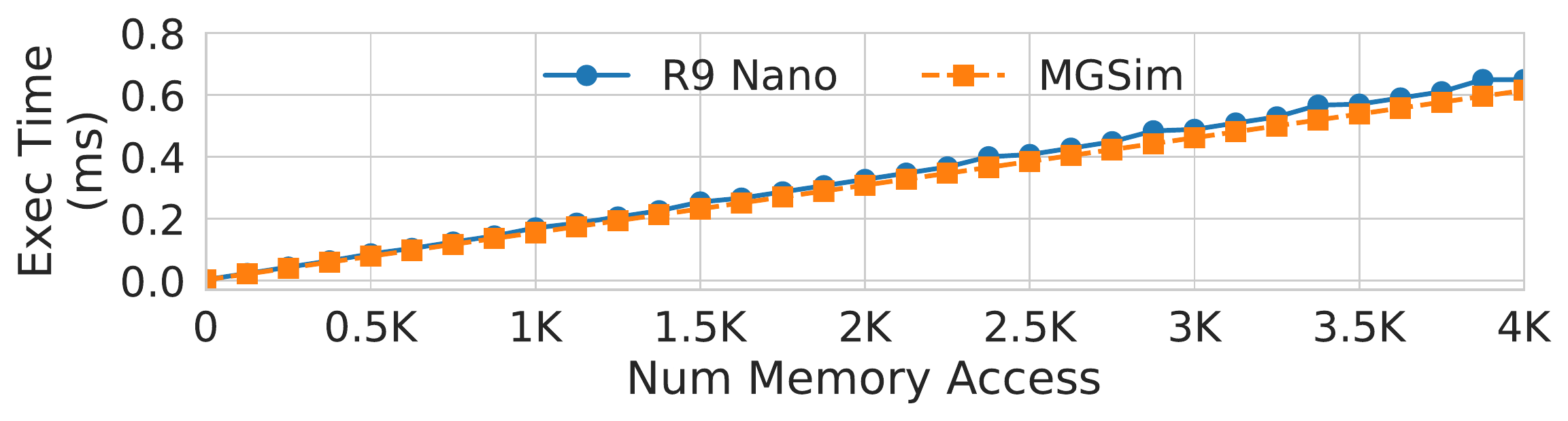}
}\\
\subfloat[L2 Read\vspace{-0.3cm}\label{fig:l2_micro_benchmark}] {
    \includegraphics[width=\columnwidth]{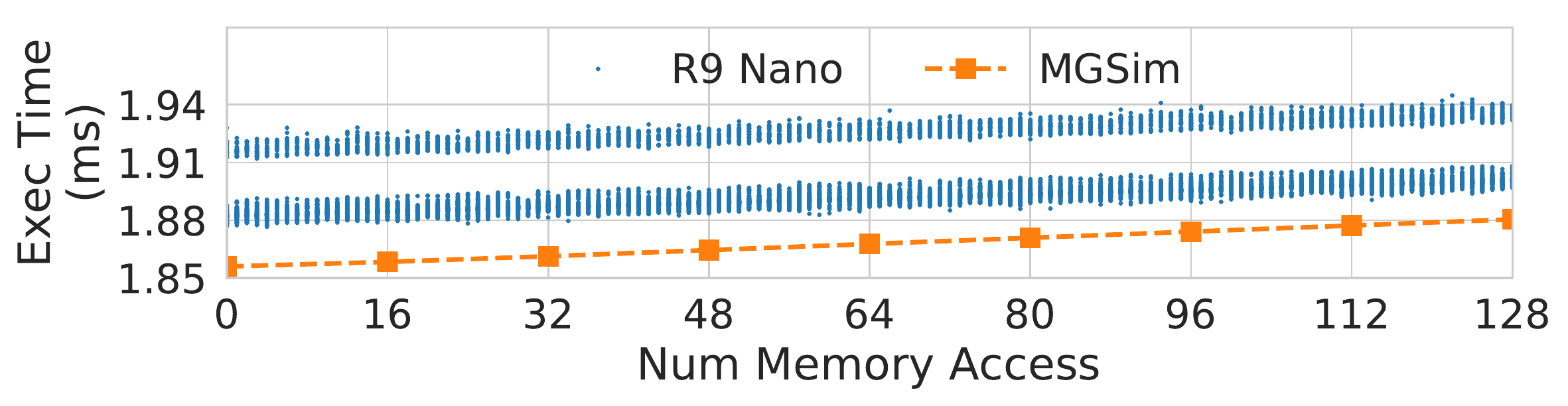}
}\\
\subfloat[DRAM Read\label{fig:dram_micro_benchmark}] {
    \includegraphics[width=\columnwidth]{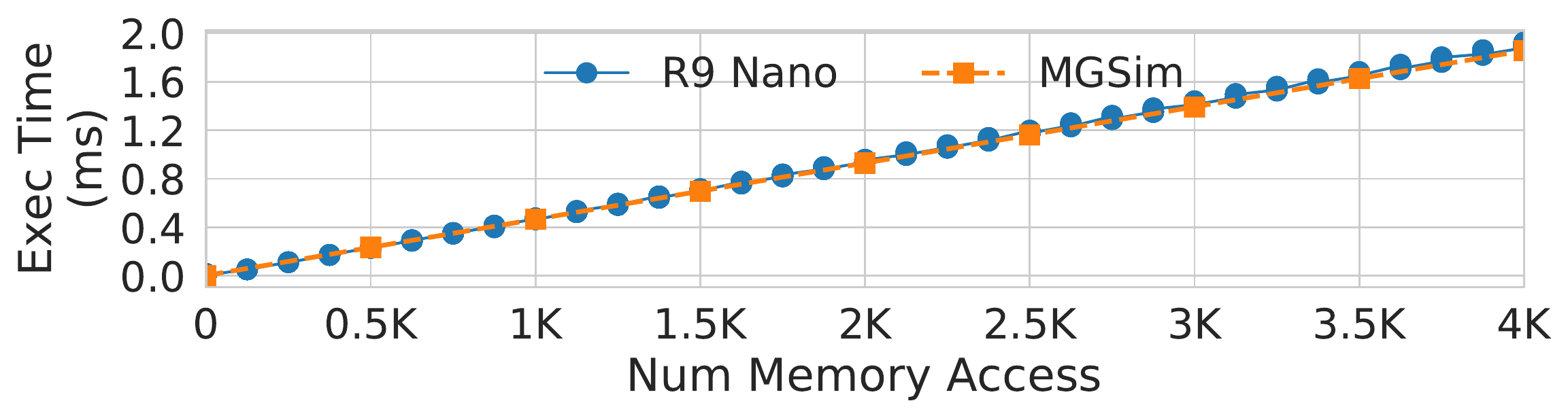}
}
\caption{Execution time comparison between R9 Nano and
MGSim on micro-benchmarks.}
\label{fig:micro_benchmarks}
\end{figure}

Figure~\ref{fig:alu_micro_benchmark} presents the execution time when using the
ALU micro-benchmark. We consider the execution time as we increase the
number of ALU operations (see X-axis).  As we can observe, the execution time
demonstrates a staircase behavior, which is the result of instruction cache
misses.  In particular, as each cache line in the CU's L1 Instruction Cache can
store 16 ALU instructions, we have 1 cache miss and 15 hits for every cache
line read.  From the slope of the time curve, 
we can conclude that the SIMD
unit takes 5 cycles to execute one instruction, and from the 
step's height, we know that the GPU spends 300+ cycles to service a cache miss.
The dashed line in the figure shows the simulator's reported execution 
time, demonstrating that our simulator can reproduce both the 
cache instruction misses and the
pipeline latencies accurately.

Figure~\ref{fig:micro_benchmarks} also explains the behavior of the memory
system of the R9 Nano GPU when using the remaining three micro-benchmarks: L1
Access, DRAM Access, and L2 Access.  Figure~\ref{fig:l1_micro_benchmark}
suggests that each L1 hit takes around 150 cycles, which is a very long time for
L1 caches.  However, if we compare these results with
Figure~\ref{fig:l2_micro_benchmark}, which shows that each L2 cache hit takes
approximately 140 - 150 cycles, we can draw the conclusion that L1V is disabled
by default in the ROCm platform. Therefore, we also disable the L1V cache in
our simulator to represent the real hardware.


As to the L2 Access micro-benchmark, since we run it on real GPU for a large
number of accesses, we use a blue dot to represent each reading in
Figure~\ref{fig:l2_micro_benchmark}.  The two groups of blue dots, separated by
a 0.03ms gap, is a result of an occasional DRAM refresh, which adds a small
amount of time to the overall execution time.  As we observe, although MGSim
underestimates the execution time, the error is very small ($1.5\%$) and, more
importantly, we can track trends that match the real GPU.

Figure~\ref{fig:dram_micro_benchmark} shows the execution time when running the
DRAM Access micro-benchmark. These results reveal that an L2 miss takes
approximately 460 cycles to service, which is the time required to traverse the
whole memory hierarchy. The combined results shown in
Figure~\ref{fig:micro_benchmarks} demonstrate that MGSim is capable of
simulating each layer of the memory hierarchy with very high accuracy.

\subsection{Simulator Validation with our \MGMark{} Suite}

\begin{figure}
    \centering
    \includegraphics[width=\columnwidth]{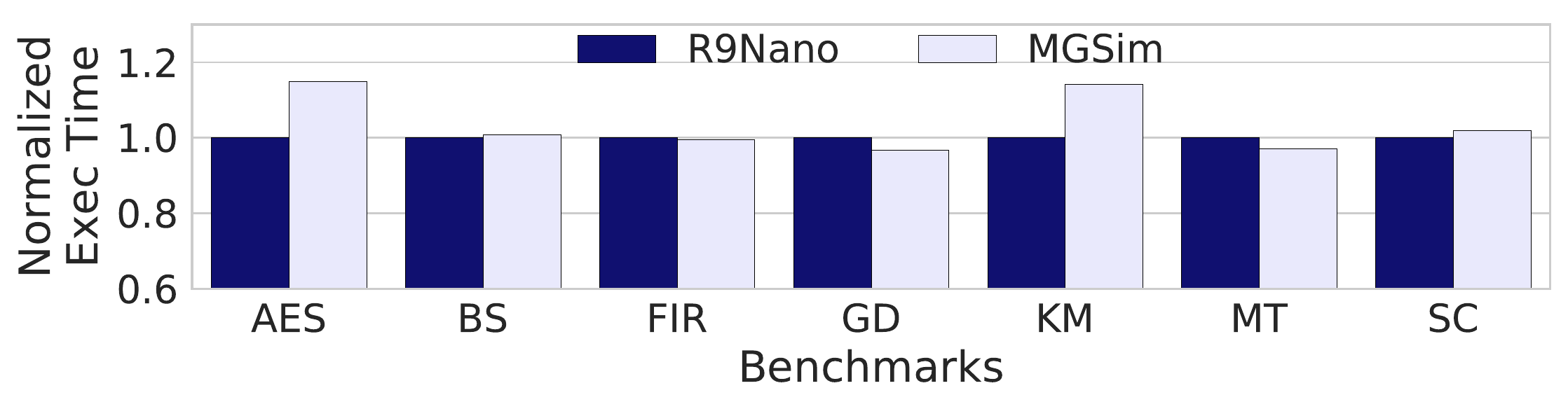}
    \caption{Execution time comparison between R9 Nano and MGSim
    for our proposed \MGMark{} suite.}
    \label{fig:validation}
\end{figure}

To validate our simulator using system workloads, we run the full set of
benchmarks included in our \MGMark{} suite.  As
we can see from Figure~\ref{fig:validation}, 
except in two cases (AES and KM), our simulator achieves almost
identical execution times. Overall, MGSim achieves performance numbers within
$5.5\%$ of the measured hardware runs.  The largest discrepancy is less than
$15\%$ in the tested benchmarks. 

\subsection{Simulator Performance}

MGSim was developed to deliver scalable simulation performance.  To
demonstrate this feature, we run MGSim configured with a single-GPU running the
MT benchmark.  Our experiment reveals that our simulator can reach 
$\approx27$ kilo-instruction per second~(KIPS) in terms of simulation speed.
Although simulators support different ISAs and model distinct components, to
put this value into perspective, we run the same experiment in two other
state-of-the-art simulators: \mbox{Multi2Sim 5.0} and GPGPUSim. We obtain
$\approx1.6$ KIPS and $\approx0.8$ KIPS, respectively. 

\begin{figure}[!t] 
\subfloat[\vspace{-0.1cm}Parallel emulation speedup\vspace{-0.2cm}]{
\includegraphics[width=\columnwidth]{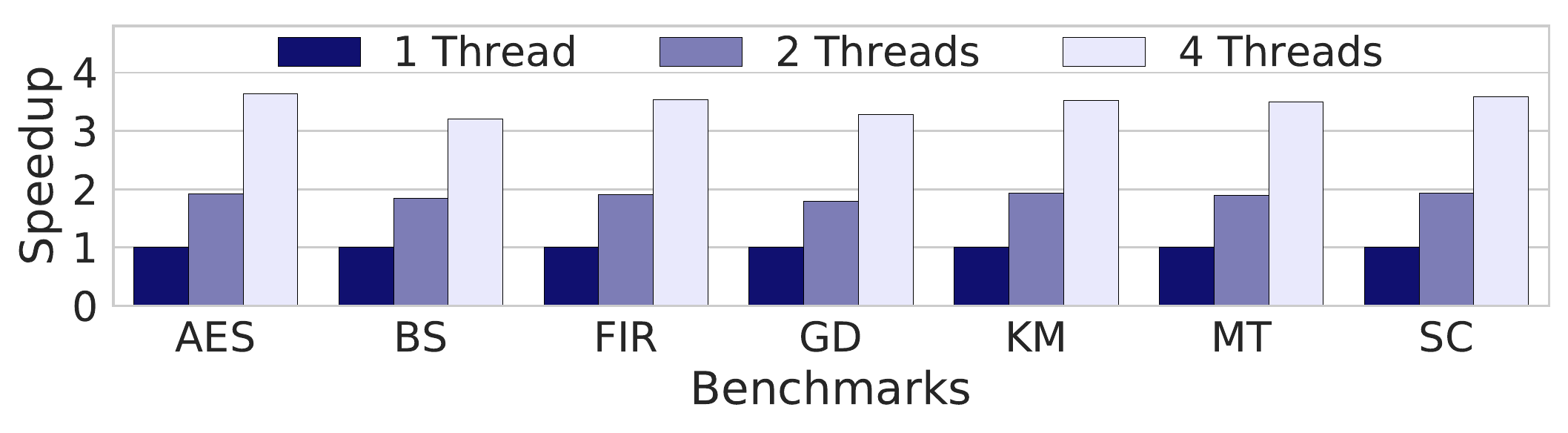} } \\
\subfloat[Parallel detailed simulation speedup]{
\includegraphics[width=\columnwidth]{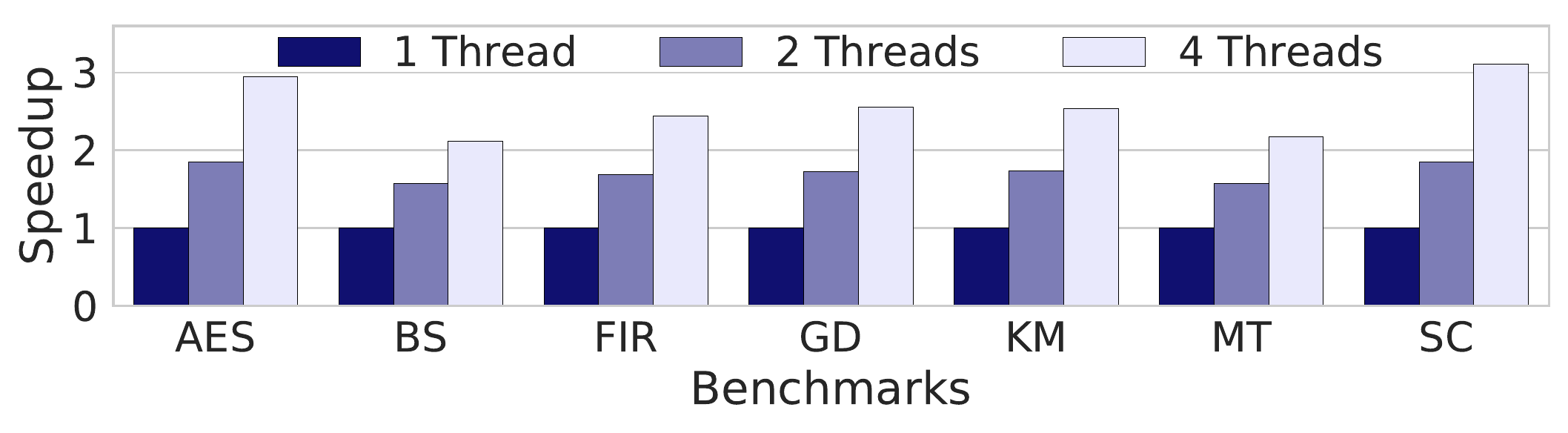} }
\caption{Multi-threaded emulation and simulation scalability.}
\label{fig:parallel} \end{figure}

To support efficient design-space exploration in the context of multi-GPU
systems, unlike contemporary GPU simulators, we designed MGSim with built-in
multi-threaded execution to further accelerate the speed of simulations. Our
simulations can take advantage of the multi-threaded/multi-core capabilities of
contemporary CPU platforms.  As shown in Figure~\ref{fig:parallel}, MGSim
achieves good scalability when using multiple threads to run simulations.
In particular, when 4 cores are used in the Intel Core i7-4770 CPU platform,
MGSim can achieve $3.5\times$ and $2.5\times$ speedups in functional emulation
and architectural simulation, respectively, while preserving the same level of
accuracy as in single-threaded simulation.  

\subsection{Evaluating Multi-GPU Configurations} \label{sec:multi-gpu-result}

\begin{figure}[!t] 
  \subfloat[Normalized execution time\vspace{-0.1cm}\label{fig:multigpu_perf}]{
    \includegraphics[width=\columnwidth]{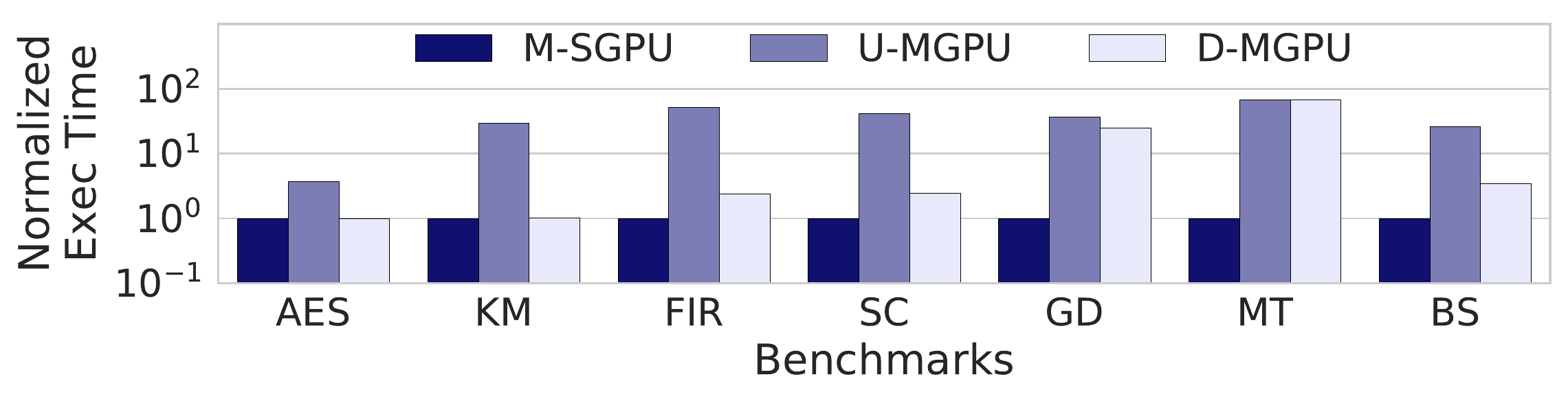}
  } \\
  \subfloat[Cross-GPU traffic\label{fig:multigpu_traffic}] {
    \includegraphics[width=\columnwidth]{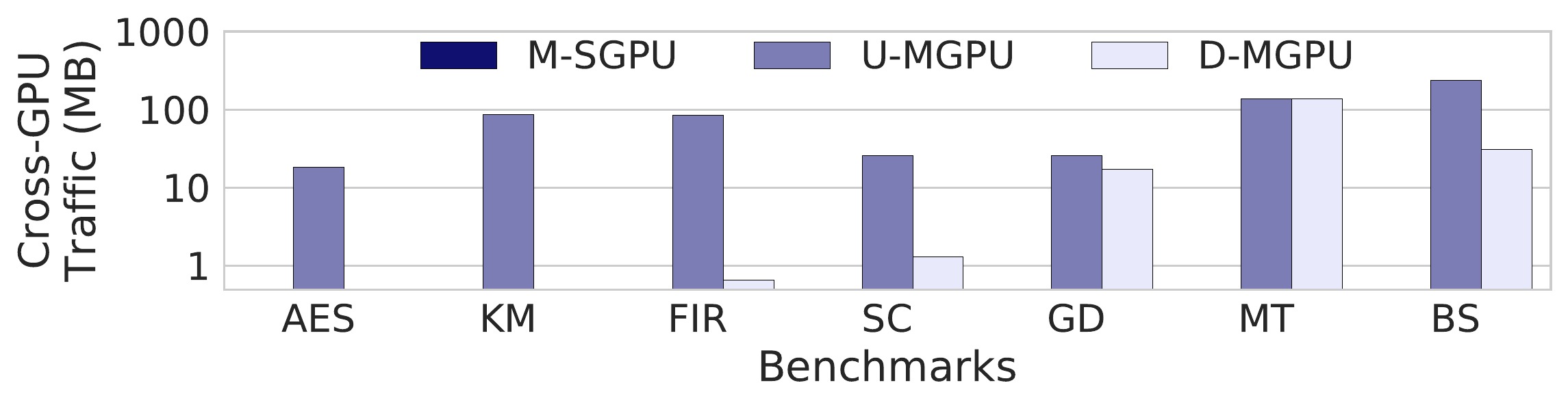}
  }
  \caption{Comparison of multi-GPU execution on different system configurations.}
  \label{fig:multigpu} 
\end{figure}

So far, we have validated MGSim considering single-GPU scenarios employing both
micro-benchmarks and our \MGMark{} suite.  Next, we evaluate the utility of
MGSim, simulating the two multi-GPU configurations defined in
Section~\ref{subsec:multigpu-confs}: \UMGPU, a unified logical GPU
configuration, which is widely adopted by the microarchitecture research
community; and \DMGPU, a discrete multi-GPU configuration. Also, to help us
analyze performance scaling, we use a baseline design that consists of a
Monolithic Single GPU~(\MSGPU) that combines 256 CUs on a single die, a GPU
that would be impractical to fabricate using today's technology.

Figure~\ref{fig:multigpu} presents the relationship between the cross-GPU
traffic and overall performance.  From the figure, we clearly see that the
cross-GPU communication is a bottleneck in the full system, as the traffic on
the interconnect is strongly correlated with the total execution time.

\UMGPU{} generally shows much larger slowdowns compared to \DMGPU\@. This
is because the programmer cannot control where the data is placed or where a
kernel is launched in \UMGPU{} design.  The lack of data-affinity scheduling
produces a large amount of cross-GPU traffic, and hence, significantly reduces
overall performance.

We also see that the different collaborative execution patterns play a roll in
overall performance.  As AES and KM follow the Partitioned Data pattern,
the programmer can eliminate all cross-GPU traffic, leading to the high performance
in \DMGPU\@.  In FIR and SC, occasional cross-GPU 
accesses occur when using an Adjacent
Access pattern, leading to a lower performance in \DMGPU{} as compared to
\MSGPU\@.  SC is worse, as compared to FIR, because it needs to load more data
from remote GPUs.  GD and MT both need to read and/or write a relatively large
amount of data from a remote GPU, suggesting Gather and Scatter are patterns
that place high demands on a multi-GPU communication.  Finally, we see \DMGPU{}
outperforms \UMGPU{} in the BS benchmarks. Although BS has an Irregular access
pattern, a majority of the swapping occurs between adjacent elements, making a
proper data partitioning still useful in improving performance.

We can draw the following design insights from the results of this case study.
1) Although unifying multiple GPUs under a single GPU interface simplifies
programmability, the performance penalty is not negligible. Future research
needs to explore solutions to reduce cross-GPU traffic to effectively
leverage an unified-GPU system.  2) Multi-GPU systems that use unified memory
and run workloads that generate cross-GPU memory accesses will require very
high bandwidth between GPUs to make a multi-GPU system scalable.  3) Multi-GPU
programmers need to have a clear picture on which collaborative execution
pattern they are adopting in order to anticipate 
cross-GPU traffic. Programmers should
also avoid patterns that generate too much cross-GPU traffic.  4) As
programmers are familiar with the programming model of discrete-GPUs, giving
back control to the programmer can be a reasonable solution to multi-GPU
systems.

\section{Related Work}

\textbf{GPU Simulators:} Ever since GPUs were introduced for 
high-performance general-purpose computing, 
researchers have developed GPU architectural
simulators to support the research community to perform architectural
exploration.  GPGPU-Sim~\cite{gpgpusim} and Multi2Sim~\cite{multi2sim} are two
of a number publicly available GPU simulators that modeled GPUs based on
NVIDIA's PTX ISA and AMD's GCN1 ISA, respectively.  The Gem5 AMD GPU
model~\cite{amdgpu} is a recent GPU simulator developed in parallel with MGSim,
and is also capable of simulating the GCN3 ISA\@.  While MGSim is inspired by
these predecessor simulators, MGSim emphasizes strong software engineering
principles, high performance parallel simulation, and multi-GPU system
modeling.

\textbf{Parallel GPU simulators:} To accelerate GPU simulation, parallel GPU
simulators have been
proposed~~\cite{barra}~\cite{gputejas}~\cite{lee2013parallel}~\cite{lee2016parallel}.
Barra~\cite{barra} mainly focuses on parallel functional emulation, which is
very different from MGSim, since it performs both emulation and timing
simulation.  GPUTejas~\cite{gputejas} is a Java-based, trace-driven, parallel
architectural simulator that can achieve high performance and scalability.
Instead of trace-driven, our simulator is execution-driven in order to support
the ``no-magic'' and ``track both timing and data'' design principles.  The
parallel simulator framework proposed by Lee et at.~\cite{lee2013parallel,
lee2016parallel} modifies GPGPUSim and only synchronizes when the processor
accesses the memory system.  Different from GPUTejas and Lee et
al.'s frameworks, we achieve scalable speedup without compromising simulation
accuracy.  We also deliver a next-generation GPU simulator that can simulate a
new ISA and multi-GPU systems.

\textbf{GPU Computing Benchmarks:} Because of the rising popularity of general
purpose computing on GPUs, a significant amount of effort has been put into
creating benchmark suites, such as Rodinia~\cite{che2009rodinia},
Parboil~\cite{stratton2012parboil} and
Lonestar~\cite{burtscher2012quantitative}.  Instead of designing for multi-GPU
systems, these benchmark suites target single GPU computing capabilities.  In
addition, Chai~\cite{gomez2017chai} and Hetero-Mark~\cite{heteromark} are
benchmark suites that specifically focus on simulating concurrent CPU-GPU
execution.  MGMark is different from existing benchmark suites, as it targets
multi-GPU systems that support unified memory and cross-GPU memory access.

\textbf{Multi-GPU Benchmarks:} Ben-Nun et al. developed
Maps-Multi~\cite{ben2015memory} and MGBench~\cite{mgbench}, 
a framework that categorizes 
multi-GPU memory access patterns and
proposes an approach that can schedule memory location and kernel execution
efficiently. The goal of our work is to provide a workload
suite to evaluate multi-GPU system with modern features, 
including unified memory and
cross-GPU memory access, which are not considered in Maps-Multi and MGBench.
Also, our benchmark suite covers a broader range of multi-GPU 
execution patterns,
compared to MGBench, which only includes two full benchmarks.

\textbf{Multi-GPU Micro-architecture Research:} More recently, the research
community has started to study how to efficiently accelerate computing with
multi-GPU systems.  As major bottlenecks of the system are the cross-GPU
interconnect and the memory system, research has focused on optimizing
memory organization.  Ziabari et al.~\cite{umh} proposed unified memory
hierarchy~(UMH) and NMOESI, using the large GPU DRAMs as cache units for 
system memory, achieving CPU-multi-GPU memory coherency.  MCM-GPU~\cite{mcmgpu}
considers a multi-chip module that encapsulates multiple GPUs in the same
package.  They introduced an L1.5 cache and used memory affinity scheduling to
reduce the cross-GPU traffic. A NUMA aware multi-GPU system, proposed by Milic
et al.~\cite{milic2017beyond}, also tries to reduce traffic on the
interconnect.  While these studies are related to our own, we do not propose a
new architecture nor algorithm, but instead, deliver a framework to explore
Multi-GPU systems and explore the possibility of giving control of
the multi-GPU system back to the programmer.

\balance
\section{Conclusion}

With the development of multi-GPU systems, the research 
community demands new tools to explore faster and scalable multi-GPU
designs.
In this paper, we have proposed a new, flexible, and high-performance,
parallel multi-GPU simulator MGSim.
We have extensively validated MGSim with both micro-benchmarks and 
full workloads against a real GPU\@.
We also describe \MGMark, a new benchmark suite for exploring
multi-GPU execution patterns. 
Together, MGSim and \MGMark{} serve as a novel 
framework that can be used to explore new and emerging multi-GPU systems.

In this paper, we presented a case study, comparing
a discrete multi-GPU system
with a unified multi-GPU system.
We draw design lessons from our case study, suggesting
that exposing a true multi-GPU interface to programmers
is a valuable solution, but requires the programmer to have 
a clear picture of the underlying program pattern.
We found that unifying the multi-GPU interface introduces a significant
amount of cross-GPU traffic, and thus, requires a high-bandwidth
interconnect as well as an efficient scheduling mechanism.

Designing a computer architecture simulator is a long-term effort. 
Despite the reasonable overall accuracy we can achieve, we will 
continue to support the simulator for the community, 
adding new features~(e.g., supporting 
atomic operations) and additional workloads. 
We also plan to explore the multi-GPU design space more thoroughly, 
including different cross-GPU network topologies, 
network fabrics, and scaling the number of GPUs in the system.

\bibliographystyle{plain}
\bibliography{references}

\end{document}